\RequirePackage{ifpdf}
\ifpdf 
\documentclass[pdftex]{sigma}
\else
\documentclass{sigma}
\fi

\begin{document}
\allowdisplaybreaks

\renewcommand{\thefootnote}{$\star$}

\renewcommand{\PaperNumber}{053}

\FirstPageHeading

\ShortArticleName{Cosmological Symmetry Breaking and Generation of Electromagnetic Field}

\ArticleName{Cosmological Symmetry Breaking\\
 and Generation of Electromagnetic Field\footnote{This
paper is a contribution to the Proceedings of the Eighth
International Conference ``Symmetry in Nonlinear Mathematical
Physics'' (June 21--27, 2009, Kyiv, Ukraine). The full collection
is available at
\href{http://www.emis.de/journals/SIGMA/symmetry2009.html}{http://www.emis.de/journals/SIGMA/symmetry2009.html}}}

\Author{Michiyasu NAGASAWA}

\AuthorNameForHeading{M. Nagasawa}

\Address{Department of Information Science, Kanagawa University, Tshuchiya 2946, Hiratsuka-shi, \\
Kanagawa-ken, 259-1293 Japan}
\Email{\href{mailto:nagasawa@info.kanagawa-u.ac.jp}{nagasawa@info.kanagawa-u.ac.jp}}

\ArticleDates{Received November 15, 2009, in f\/inal form May 14, 2010;  Published online June 29, 2010}

\Abstract{Cosmological phase transitions accompanied by some kind of symmetry breaking
would cause the creation of topological defects and the resulting production of primordial
magnetic f\/ield. Moreover, such a procedure inevitably af\/fects the cosmic background ra\-dia\-tion
and it may be observed today. Motivated by the existence of stabilized embedded defects
in the standard model of elementary interactions, we discuss their application to
the cosmological electromagnetic f\/ield generation.}

\Keywords{cosmology; defect; pion string; magnetic f\/ield}

\Classification{83F05}

\section{Introduction}

Particle cosmology predicts that our universe has experienced some phase transitions
associated with the unif\/ication theory of the interactions. One of the important
applications of such a~transition to the cosmic evolution would be the generation of
topological defects~\cite{KVS}. These defects are topologically stable and whether their
formation is possible or not and the type of their structure can be judged by the way of
symmetry breaking at the phase transition. In addition to conventional topologically stable
defects, another kind of cosmological defects are suggested as an interesting product in
the early universe~\cite{embstr}. Even when the topologically charged  conf\/iguration is
not to be prevented from becoming a trivial one, such a f\/ield conf\/iguration can be constructed
so that it satisf\/ies the equation of motion although it is not energetically preferable.
That is, embedded defects are unstable at zero temperature. In the early universe, however,
the f\/inite temperature plasma existed so that they can be stabilized because of the asymmetry
between charged and neutral scalar components~\cite{NB1}. At the low temperature, the photon decoupling
occurs on the other hand. Then defects undergo core phase transition and/or decay, which
might bring primordial magnetic f\/ield generation. Although at present there exist magnetic f\/ields
in various astrophysical scales, the origin of galactic magnetic f\/ields $\sim 10^{-6}$~G is
not revealed completely~\cite{KZ}. We will see pion strings can provide seed magnetic f\/ields which
evolve to present large scale f\/ields. Note that, in addition, some kinds of ef\/fects on cosmic
microwave background radiation could be observed by the interaction between pion f\/ields and cosmic
background photons.

A cosmic string is a two-dimensional defect and it is widely investigated since it has various
cosmological signif\/icances. If the ef\/fective potential has the form of a Mexican hat,
a phase transition accompanied by some kind of symmetry breaking occurs and cosmic strings
are produced. For example, suppose a complex scalar f\/iled $\phi$ obeys
\begin{gather*}
V(\phi)=\frac{\lambda}{4}\big(\phi^*\phi -\eta^2\big)^2 ,
\end{gather*}
where $\eta$ is the energy scale of the symmetry breaking.
At the region suf\/f\/iciently far from the string core, the string conf\/iguration can be expressed as
\begin{gather*}
\phi =\eta e^{in\theta} ,
\end{gather*}
where $n$ is a winding number.
The linear mass density of the string, $\mu$, would be
\begin{gather}
\mu \approx \eta^2 ,
\qquad \mbox{and}\qquad
G\mu =\left( \frac{\eta}{M_{\rm Pl}}\right)^2 \label{eq:gmu}
\end{gather}
is the parameter which determines the magnitude of the string ef\/fect where $M_{\rm Pl}$ is
the Planck mass.

Even in the cases when the vacuum structure of a particle physics model is trivial, a string
solution does exist. Although the conf\/iguration of embedded defects satisf\/ies equations of motion,
they are topologically, and in general also dynamically unstable. If three real scalar f\/ield
components, $\phi =(\phi_1,\phi_2,\phi_3)$, obeys the ef\/fective potential as
\begin{gather*}
V(\phi)=\frac{\lambda}{4}\left(\sum_{i=1}^3\phi_i^2 -\eta^2\right)^2 ,
\end{gather*}
then no string would be produced since the vacuum manifold is $S^2$ as
\begin{gather*}
\sum_{i=1}^3\phi_i^2=\eta^2 .
\end{gather*}
However, by freezing out certain components such as $\phi_3=0$, the vacuum structure is modif\/ied to
\begin{gather*}
\sum_{i=1}^2\phi_i^2=\eta^2 ,
\end{gather*}
which is equal to $S^1$ and a string conf\/iguration can be realized.

In the next section, production and evolution of pion strings are brief\/ly introduced. It is indicated that
the pion string can be superconducting which means cosmic vortons may be formed and/or long horizon-scale
strings can be observed as line-like objects in contrast to usual particle-like ones. Then in Section~\ref{section3}
we consider another possibility of electromagnetic interaction provided by pion strings, that is, the rotation
of polarization axis of light, which should inevitably af\/fect the microwave background radiation anisotropy
also as a line-like signature. Moreover, the magnetic f\/ield generation by pion strings will be discussed
in Section~\ref{section4}.

\section{Pion string}\label{section2}

One example of embedded global string is a pion string~\cite{ZHB}. It is predicted in the context of
the standard model of strong interaction and produced at the QCD phase transition.
Below the conf\/inement scale, this model is described by a sigma model as{\samepage
\begin{gather}
{\cal L}_0=\frac{1}{2}\partial_\mu\sigma \partial^\mu\sigma +\frac{1}{2}\partial_\mu\vec{\pi}
\partial^\mu\vec{\pi}-V_0 ,\qquad V_0\equiv \frac{\lambda}{4}\big(\sigma^2+\vec{\pi}^2-\eta^2\big)^2 ,
\label{eq:lag}
\end{gather}
involving the sigma f\/ield $\sigma$ and the three pions $\vec{\pi}=(\pi^0,\pi^1,\pi^2)$.}

Here we brief\/ly introduce how f\/inite temperature ef\/fect stabilizes pion strings~\cite{NB2}.
When the background photon plasma in the early universe can be regarded as a thermal bath,
the interaction between charged f\/ields and photon could be included into the ef\/fective
potential~\cite{NB1}~as
\begin{gather}
V_T=V_0+\frac{1}{2}e^2\kappa T^2\big\{ \big(\pi^1\big)^2+\big(\pi^2\big)^2\big\} ,
\end{gather}
where $T$ is the temperature of the universe, $e$ is the elementary
electric charge and $\kappa \sim {\cal O}(1)$ is a constant. Hence in
contrast to the neutral f\/ield that still obeys a double-well type potential, the charged f\/ield is
compelled to stay on the symmetric state. The destabilization temperature can be calculated~\cite{NB1} as
\begin{gather*}
T_D=2\lambda^{1/2}\kappa^{-1/2}e^{-1}\eta ,
\end{gather*}
by analyzing the stability of the pion string solution using a variational method for an unstable
mode of the cylindrically symmetric conf\/iguration. At the destabilization temperature, some kinds of
embedded defects undergo a core phase transition~\cite{AP}. When the f\/inite temperature ef\/fect dominates,
the neutral f\/ields have a string conf\/iguration and the charged ones are frozen out so that
$\sqrt{\left(\pi^1\right)^2+\left(\pi^2\right)^2}=0$ everywhere. In contrast to that, after a core phase
transition, the scalar f\/ield has a f\/inite expectation value also at the string core, that is, it settles
in the ground state in the whole universe. However, the neutral f\/ield conf\/iguration is not destroyed.
The charged f\/ields have a f\/inite expectation value only at the string core in order to compensate the neutral
f\/ields as $\sqrt{\sigma^2+\left(\vec{\pi}\right)^2}=\eta$. On the contrary, the charged f\/ields keep zero
expectation value far from the string core since the winding number is a kind of topological charge and
it must be conserved. In the pion string case, it has been shown a core phase transition should occur
by the results of numerical simulations~\cite{NB2} in which pion strings are formed at $T\sim\eta$
and $T_D=0.04\eta-0.05\eta$.

For the pion string, since
\begin{gather*}
\eta =\Lambda_{\rm QCD}\approx 200~\mbox{MeV} ,
\end{gather*}
gravitational ef\/fect should be negligible during the cosmic evolution as you can see in
the equation~(\ref{eq:gmu}). Thus the most distinct evidence of pion strings could be given by
the electromagnetic interaction. Note that the pion strings are expected to be formed in LHC Pb--Pb
collision experiments following the Kibble--Zurek mechanism, although it is dif\/ferent from
the cosmological scenario. These ef\/fects could be observable and bring distinction compared to
conventional predictions~\cite{MLNZH}.

One promising electromagnetic ef\/fect by pion strings comes from the superconductivity~\cite{Witt}.
After the core phase transition, charged f\/ields have f\/inite expectation value as
\begin{gather*}
\sqrt{(\pi^1)^2+(\pi^2)^2}=\phi_c(x,y)e^{i\varphi_c\left( z,t\right)} ,
\end{gather*}
where $xy$-plain lies perpendicular to the string and the phase has a spatial gradient along the string
so that the electric current will be generated whose amplitude, $J_c$, will be
\begin{gather}
J_c\sim e\frac{d\varphi_c}{dz} .
\end{gather}
The evidence of the existence of superconducting strings can be found by cosmological observa\-tions~\cite{OTWV}.
In addition, superconducting sting loops may become vortons~\cite{DS} in some cases and have potential
cosmological consequences~\cite{BCDT}. However, since the linear mass density is small compared to, for example,
GUT scale strings, the particle emission and cosmic microwave radiation distortion caused by superconducting
pion strings will be overcome by other ef\/fects. Thus the possible detection of a string might be obtained by
the spatial distribution pattern of emission or distortion. This is very similar to the case of a gravitational
lensing~\cite{HV}. Although the lensing ef\/fect itself occurs frequently in the universe, a line-like spatial
distribution feature must be characteristic. Also in the case of pion strings, inf\/initely long strings would
show a~distinct pattern compared to other various f\/luctuation sources and could directly prove the existence
of horizon scale long strings.

Here we do not discuss the observational consequences in detail but just refer the result of 3-dimensional
numerical simulations. Their results have conf\/irmed that the pion string has a superconducting mode after
it experiences a core phase transition. We have simulated the evolution of a pion string using a code based
on the one employed in~\cite{NB2}. In calculations, all dimensional quantities are rescaled by appropriate
powers of $\eta$ to make them dimensionless. The spatial resolution is $\Delta x = 0.5\eta^{-1}$, and
the time steps are chosen as $\Delta t=\frac{1}{10}\Delta x$. The size of box is $300^2\times 600$ with
periodic boundary in the $z$ direction. For $x$ and $y$ directions, Neumann boundary conditions are employed
but we have checked that the result is insensitive to this particular choice. First we set up the initial
conf\/iguration as an inf\/initely long straight global string with a width, $100\eta^{-1}$, along the $z$-axis
which is formed by the neutral components of the scalar f\/ields $\sigma$ and $\pi_0$ whose center resides
in the midpoint of the $xy$-plane. For charged components, $\pi_1 =\pi_2 =0$. Although such a highly symmetric
conf\/iguration might be too ideal, it would be appropriate to see whether the superconducting can occur or not
for the horizon scale string. We then add thermal energy to the conf\/iguration in the form of kinetic energy,
that is, the time derivative of the scalar f\/ields. Its amplitude is $0.1 \times T^2$, and the allocation to
the four components of the scalar f\/ield is chosen at random. The initial temperature is well below $T_D$, and
equal to $0.01\eta$. Then, four scalar f\/ields $\sigma$ and $\vec{\pi}$ are evolved numerically on a~three-dimensional lattice by means of the equations of motion derived from~(\ref{eq:lag}). During each simulation,
the background temperature is constant and the cosmic expansion is not taken into account. The results of
20 times simulations with dif\/ferent initial settings of the phases of the time derivative of the scalar f\/ields
show that the distribution of $\varphi_c$ shows the winding number appears in some cases. The probability of
winding number appearance has been derived to be 20\% which coincides with the analytical estimation as about
$16$\% based on the random phase distribution of $600\eta^{-1}/T^{-1}$ steps along the string. Thus it is
conf\/irmed that the pion string after the core phase transition can have superconductivity.
Note that we have calculated for various patterns of the numerical values of parameters and conf\/irmed they
do not af\/fect the essential results.

\section{Interaction of scalar f\/ield with electromagnetic f\/ield}\label{section3}

Here we deal with one of the possibilities of observing the line-like signature by horizon scale pion strings.
In general, the following type of interaction ${\cal L}_{\rm int}$ between the electromagnetic f\/iled and
a certain kind of f\/ield, ${\cal O}_\mu$, would appear in the Lagrangian when the anomaly or the Chern--Simons term
is taken into account~\cite{CF}. For example,
\begin{gather*}
{\cal L}_{\rm int} =-\frac{1}{2}{\cal O}_\mu A_\nu\varepsilon^{\mu\nu\alpha\beta}F_{\alpha\beta},
\end{gather*}
where $A_\mu$ is the electromagnetic f\/ield and $F_{\mu\nu}$ is its strength. Then the equations of motion
for electromagnetic f\/ield should be modif\/ied as
\begin{gather*}
\partial_\mu F^{\mu\nu}=4\pi J^\nu +{\cal O}_\mu\varepsilon^{\mu\nu\alpha\beta}F_{\alpha\beta},
\end{gather*}
so that the polarization of background radiation would occur because of the f\/ield, ${\cal O}_\mu$.

In case of the pion string, there exists an interaction between the pion f\/ield $\pi^0$ and the electromagnetic
f\/ield as
\begin{gather}
{\cal L}_{\rm int} =-\frac{N_c\alpha}{24\pi}\frac{\pi^0}{\eta}\varepsilon^{\mu\nu\alpha\beta}F_{\mu\nu}
F_{\alpha\beta} ,\label{eq:Lint}
\end{gather}
where $N_c=3$ and $\alpha$ is a f\/ine structure constant~\cite{Pisa}. We can estimate the ef\/fect on the light
propagation based on modif\/ied Maxwell equations using this Lagrangian.

Since we would like to know how the propagation of the cosmic background radiation can be af\/fected by pion
strings, we assume the pion string f\/ield conf\/iguration is not disturbed by electromagnetic f\/ield and pion
f\/ields contribute just as background current sources. Moreover, the time evolution of string distribution
is neglected so that the string spatial position can be regarded as f\/ixed during the photon propagation.
Such assumptions can be applied when the string motion is suf\/f\/iciently slow. Then the equations without
an ordinary electromagnetic current under the string background will be written as
\begin{gather*}
  \nabla {\bf E}=-{\bf \Phi}\cdot{\bf B} ,\qquad
-\frac{\partial {\bf E}}{\partial t}+\nabla\times{\bf B}={\bf \Phi}\times{\bf E} ,\qquad
 {\bf \Phi}\equiv\frac{N_c\alpha}{3\pi}\nabla\frac{\pi^0}{\eta} ,
\end{gather*}
and the dispersion relation becomes
\begin{gather*}
k^2=\omega^2\pm\omega\Phi\cos\varphi\left(1-\frac{\Phi^2\sin^2\varphi}{\omega^2-k^2}\right)^{-1/2} ,
\end{gather*}
where $\Phi\equiv |{\bf \Phi}|$ and $\varphi$ is the angle between ${\bf \Phi}$ and the wave number
vector ${\bf k}$. Under the approximation that $\Phi$ is suf\/f\/iciently small,
\begin{gather*}
k\cong \omega\pm\frac{\Phi}{2}\cos\varphi .
\end{gather*}
Thus the rotation angle dif\/ference of polarization axis between left-handed polarization and right-handed
one, $\Delta\psi$, can be estimated dependent on the direction of the string axis to the line of sight as
\begin{gather*}
\Delta\psi =\frac{\Phi}{2}\cos\varphi d=\frac{N_c\alpha}{6\pi}\left|\nabla\frac{\pi^0}{\eta}\right|d\cos\varphi   ,
\end{gather*}
where $d$ is the distance a photon travels from the source to the observer. When the light travels
perpendicular to the string and suf\/f\/iciently close to the string core where $\pi_0 =0$,
$\big|\nabla\frac{\pi^0}{\eta}\big|d\sim 1$ if~$d$ is suf\/f\/iciently large since $-\eta\le\pi_0\le\eta$
far away from the string. Then
\begin{gather}
\Delta\psi \approx 10^{-3} ,\label{eq:angle}
\end{gather}
which is the maximum value since $\cos \varphi =1$. Note that the sign depends on which side the light
passes. On the other hand, when the light travels in parallel with the string,
\begin{gather*}
\Delta\psi\approx 0 .
\end{gather*}
In any cases, the distance between the string and the light path is not so signif\/icant.

It has been shown that similarly to other defects~\cite{MSCF} pion strings also cause the polarization
rotation angle dif\/ference. What is new in the present work is the application to the pion string and
the quantitative estimation in (\ref{eq:angle}). Its amplitude is very small and hard to be detected.
Hence the string signature could be claimed when the polarization rotation angle has a line-like spatial
pattern, which is the same situation as one in the last section.

\section{Generation of magnetic f\/ield helicity from pion strings}\label{section4}

In this section,  we consider how the same interaction as (\ref{eq:Lint}) af\/fects the generation of
magnetic f\/ield. Due to this additional interaction zero mode current appears within the string core
and the azimuthal magnetic f\/ield as
\begin{gather}
B_\theta =-N_c\frac{en_c}{2\pi}\delta_s^{-\alpha/\pi}r^{-1+\alpha/\pi} ,\label{eq:Bfield}
\end{gather}
is produced\cite{BZ} where $n_c$ is a line number density of charge carriers on the string, $\delta_s$
is the string core length and $r$ is the distance from the string. This formula was derived~\cite{KM} by
solving the classical equation of motion for electromagnetic f\/ields with a source term added by
the interaction with a scalar f\/ield~(\ref{eq:Lint}). The solution was calculated under the condition
that the background scalar f\/ield has a conf\/iguration obtained from an inf\/initely long straight pion
string along the $z$-axis. The magnetic f\/ield strength generated by a pion string at the recombination
epoch has been derived~\cite{BZ} by substituting appropriate values as
\begin{gather}
B_\theta\approx 10^{-23}~\mbox{G}\ \left(\frac{r_i}{\delta_s}\right)^{\alpha/\pi}
\frac{1~\mbox{kpc}}{r_p},\label{eq:Btheta}
\end{gather}
where $r_i$ is $r$ measured at the string formation and $r_p$ is $r$ of the present scale. Although this
numerical value might be suf\/f\/icient for the seed magnetic f\/ield amplitude, we will see it can be enhanced
when the helicity conservation is taken into account.

If the twist and tangle of strings are biased when the CP violation exists, then the helicity of
magnetic f\/ield is also biased so that its conservation leads to the generation of larger magnetic f\/ield
amplitude~\cite{Vach}. First let us estimate the magnetic helicity density, ${\cal H}$, produced by pion strings.
Since it can be written as
\begin{gather*}
{\cal H}=\frac{1}{V}\int_V d^3x A_\mu B^\mu ,
\end{gather*}
within a certain volume $V$ and the vector potential can be written by the same calculation as one
in~(\ref{eq:Bfield}) as
\begin{gather*}
A_z =\frac{\pi}{\alpha}N_c\frac{en_c}{2\pi}\delta_s^{-\alpha/\pi}r^{\alpha/\pi} ,
\end{gather*}
we can calculate ${\cal H}$ if we know the complete conf\/iguration of all the pion string within $V$.
However, it would be too dif\/f\/icult to obtain such a distribution so that only a simple estimation is performed
in the following.

Using the helicity amplitude per unit string length realized by a pair of pion strings, ${\cal H}_1$, the total
helicity can be written as
\begin{gather*}
{\cal H}\simeq \epsilon_{\rm CP}n_sl_s2{\cal H}_1 ,
\end{gather*}
where $\epsilon_{\rm CP}$ is the CP violation strength which determines how much the twist and tangle of pion strings
should be biased, $n_s$ is the string number density and $l_s$ is the typical string length. In order to derive
${\cal H}_1$, here we employ a very rough estimation as
\begin{gather}
{\cal H}_1\sim B(\delta_s)A(L_s)\delta_s^2 ,\label{eq:H1}
\end{gather}
where $L_s$ is the mean separation length between pion strings. This estimation is the most optimistic one which
corresponds to the case when pion strings are so arranged that $A$ f\/ield and~$B$ f\/ield are parallel. Then
\begin{gather*}
{\cal H}\sim -\epsilon_{\rm CP}n_s\frac{N_c^2e^2}{2\pi\alpha}n_c^2l_s\delta_s\left( \frac{L_s}{\delta_s}\right)^{\alpha/\pi} .
\end{gather*}
It would be reasonable if we assume
\begin{gather*}
\delta_s\sim\Lambda_{\rm QCD}^{-1} ,
\end{gather*}
and
\begin{gather*}
L_s\sim l_s\sim n_s^{-1/3} .
\end{gather*}
Here we introduce a parameter $\beta$ as
\begin{gather*}
L_s=\beta T^{-1} ,
\end{gather*}
then the helicity density can be expressed as
\begin{gather*}
{\cal H}\sim -\epsilon_{\rm CP}2N_c^2\beta^{-2+\alpha/\pi}\left( \frac{\Lambda_{\rm QCD}}{T}\right)^{1+\alpha/\pi}T^3 ,
\end{gather*}
where $n_c\sim \Lambda_{\rm QCD}$ has been substituted. If we normalize this value by the cosmic entropy density,~$s$,
\begin{gather*}
\frac{{\cal H}}{s}\sim -\epsilon_{\rm CP}\frac{45 N_c^2}{\pi^2{\cal N}}
\beta^{-2+\alpha/\pi}\left( \frac{\Lambda_{\rm QCD}}{T}\right)^{1+\alpha/\pi} ,
\end{gather*}
where ${\cal N}$ is the number of gauge freedom at $T$. The most ef\/f\/icient case is realized when $\beta \sim 1$
and such a situation would be probable at the string formation epoch, that is, $T\sim \Lambda_{\rm QCD}$.
On the other hand, at much lower temperature, the coherence scale of the string distribution may be larger so that
the resulting helicity density would be insignif\/icant. For example, at the temperature scale $\sim 1$~MeV,
$\beta$ could be $10^{-21}$ if about one string exists within the horizon scale.

Next we have to consider the evolution of the magnetic f\/ield and whether the magnetic helicity would be conserved
or not. At the formation of the magnetic f\/ield, the magnetic Reynolds number should be suf\/f\/iciently large since
the condition as
\begin{gather*}
v\gg \frac{e^2}{4\pi} ,
\end{gather*}
seems to be satisf\/ied where $v$ is the characteristic f\/luid velocity. Thus the generated magnetic helicity should
not be erased and the following evolution would be worth to be analyzed. There exist two important scales in this
stage. One is the frozen-in scale, $L_f$, above which the helicity is conserved. It can be calculated by
the condition that the ef\/fect of the dissipation by the surrounding plasma in the classical f\/ield equation of
motion for the magnetic f\/ield is negligible compared to the damping due to the cosmological expansion as
\begin{gather*}
L_f\sim \sqrt{\frac{t}{4\pi \sigma_c}} ,
\end{gather*}
where $t$ is the cosmic time and $\sigma_c\sim T/e^2$ is the electrical conductivity of the plasma. Magnetic f\/ields
whose coherent scales are larger than $L_f$ are not dissipated and are frozen in. The other scale is the magnetic
coherence scale, $L(t)$. In one expansion time after the magnetic f\/ield generation, $L$, can be written as
\begin{gather*}
L(t_d)\sim L_i\left( \frac{t}{t_i}\right)^\xi ,
\end{gather*}
where $L_i$ is the initial coherence scale, $t_i \sim L_i/v$ is the time scale of magnetohydrodynamic turbulence
and the subscript $d$ depicts the decay of the pion string. It is assumed that $L$ grows as $\propto t^{\xi}$.
Although the value of $\xi$ is not completely determined, it is of the order of unity. For example, $\xi =1/2$
was derived in some of MHD simulation results~\cite{MBCHB} and $\xi =2/3$ was the case when the magnetic energy
density is scale free~\cite{Olse}. These two values are employed as typical ones in the following estimation.
Although these particular choices are important when the helicity conservation condition is examined, they
have little ef\/fect on the quantitative estimation of the resulting magnetic f\/ield strength. We take~$L_i$ as
a simple thermal one, $L_i\sim 1/T_d$, and the f\/luid is relativistic, $v\sim 1$, then
\begin{gather}
L(t_d)\sim \frac{1}{T_d}\left( \frac{T_{pl}}{T_d}\right)^\xi , \label{eq:Ltd}
\end{gather}
where $T_{pl}\sim 10^{19}$ GeV is the Planck temperature. When $T_d=100$~MeV,
\begin{gather*}
L_f\sim 10^9T_d^{-1} ,
\qquad \mbox{and}\qquad L(t_d)\sim 10^{10}T_d^{-1} ,
\end{gather*}
for the case $\xi =1/2$,
\begin{gather*}
L(t_d)\sim 2\times 10^{13}T_d^{-1} ,
\end{gather*}
for the case $\xi =2/3$.
When $T_d=1$ MeV,
\begin{gather*}
L_f\sim 10^{10}T_d^{-1} ,
\qquad \mbox{and}\qquad
L(t_d)\sim 10^{11}T_d^{-1} ,
\end{gather*}
for the case $\xi =1/2$,
\begin{gather*}
L(t_d)\sim 5\times 10^{14}T_d^{-1},
\end{gather*}
for the case $\xi =2/3$. In both cases,
\begin{gather*}
L_f<L(t_d),
\end{gather*}
which means the helicity conservation condition is satisf\/ied. This condition will be also satisf\/ied
in the following evolution since
\begin{gather*}
L_f\propto \left( \frac{T_d}{T}\right)^{3/2} ,
\qquad \mbox{and}\qquad L(t)=L(t_d)\left( \frac{T_d}{T}\right)^{1+\xi} .
\end{gather*}
At $T\sim 0.1$ MeV, the electron pair annihilation occurs and the electric conductivity decreases suddenly as
\begin{gather*}
\sigma_c \sim 10^{-10}\frac{m_e}{e^2} ,
\end{gather*}
where $m_e$ is the electron mass so that the frozen-in scale exceeds the coherence scale as
\begin{gather*}
\frac{L(t)}{L_f}\sim 2\times 10^{-4} ,
\end{gather*}
when $\xi =1/2$ independent of the value of $T_d$. On the other hand, when $\xi =2/3$, $L(t)$ is comparable
to $L_f$ and this is more interesting case.

Hence we can calculate the coherence scale at the recombination when $T\sim 0.1$~eV using the formula as
\begin{gather}
L_{\rm rec}=L(t_d)\left( \frac{T_d}{T_{\rm eq}}\right)^{1+\xi}\left( \frac{T_{\rm eq}}
{T_{\rm rec}}\right)^{1+\xi/2} , \label{eq:Lrec}
\end{gather}
where $T_{\rm eq}\sim 1$~eV is the matter-radiation equality time.
Then $L_{\rm rec}$ can be calculated as
\begin{gather*}
L_{\rm rec}\sim 10^{15}~{\rm cm} .
\end{gather*}
Note that although it is $10^{14}$ cm in the paper \cite{Vach}, it comes simply from the dif\/ferent treatment
of numerical factors since from the equations~(\ref{eq:Ltd}) and~(\ref{eq:Lrec}) we can see that $L_{\rm rec}$ does
not depend on $T_d$. Finally we show the numerical value of the magnetic f\/ield strength for the extreme case
when ${\cal H}/s\sim 1$ as
\begin{gather*}
B\sim 10^{-9}\ {\rm G} ,\qquad {\rm on}\ \sim 1~{\rm pc} ,
\end{gather*}
at the recombination epoch. This value is remarkably larger than that in (\ref{eq:Btheta}) and much more promising
for the explanation of the origin of the astrophysical magnetic f\/ield. Although we employ the optimistic estimation~(\ref{eq:H1}), the actual value cannot be smaller than $10^{-10}$ G if we consider the average over randomly
oriented~$A$ and~$B$ f\/ields. However, its coherence scale is smaller compared to the galactic or intergalactic scales.
In order to obtain the f\/ield amplitude on larger scales, we have to determine the detailed initial distribution of
magnetic helicity, that is, the twist and tangle of pion strings. Further investigation would be interesting.

\section{Summary}\label{section5}

In this paper, the magnetic f\/ield generation by pion strings produced at the QCD phase transition and their
possible detection by astronomical observation are considered.

For the latter purpose, the superconductivity of pion strings are demonstrated. In addition, their interaction
with the cosmic background radiation is investigated and it is shown that the rotation of light polarization
axis is caused by the string f\/ield. In both cases, the line-like structure of strings itself is essential and
the spatial distribution of microwave background distortion or rotation angle dif\/ference of polarization traces
the arrangement of the strings in our universe.

Since the observational consequences are just suggestions, our main conclusion is that pion strings can generate
primordial magnetic f\/ields which may be seeds of present cosmic magnetic f\/ields. Its amplitude can exceed
$B\sim 10^{-10}$ G and this is much higher than that in the preceding work~\cite{BZ}. Such an enhancement becomes possible by the consideration of the helicity conservation which has been shown to be reasonable for the pion string.

\subsection*{Acknowledgements}

The numerical calculations were carried out on SX8 at YITP in Kyoto University.

\pdfbookmark[1]{References}{ref}
\LastPageEnding

\end{document}